\documentclass[aps,prb,twocolumn,superscriptaddress,color,pdflatex]{revtex4}

\usepackage{graphicx}
\usepackage{dcolumn}
\usepackage{amsmath,amssymb,bbold,bm,epsfig}

\begin{document}

\newcommand{\bk}{{\bf k}}
\newcommand{\bp}{{\bf p}}
\newcommand{\bv}{{\bf v}}
\newcommand{\bq}{{\bf q}}
\newcommand{\tbq}{\tilde{\bf q}}
\newcommand{\tq}{\tilde{q}}
\newcommand{\bQ}{{\bf Q}}
\newcommand{\br}{{\bf r}}
\newcommand{\bR}{{\bf R}}
\newcommand{\bB}{{\bf B}}
\newcommand{\bA}{{\bf A}}
\newcommand{\bE}{{\bf E}}
\newcommand{\bj}{{\bf j}}
\newcommand{\bK}{{\bf K}}
\newcommand{\cS}{{\cal S}}
\newcommand{\vd}{{v_\Delta}}
\newcommand{\tr}{{\rm Tr}}
\def\id{\mathbb{1}}

\title{Topological Anderson Insulator in Three Dimensions}

\author{H.-M. Guo}
\affiliation{Department of Physics and Astronomy,
University of British Columbia, Vancouver, BC, Canada V6T 1Z1}
\affiliation{Department of Physics, Capital Normal University, Beijing 100048, China} 
\author{G. Rosenberg}
\affiliation{Department of Physics and Astronomy,
University of British Columbia, Vancouver, BC, Canada V6T 1Z1}
\author{G. Refael}
\affiliation{Department of Physics,
California Institute of Technology, 
Pasadena, CA 91125 }
\author{M. Franz}
\affiliation{Department of Physics and Astronomy,
University of British Columbia, Vancouver, BC, Canada V6T 1Z1}

\date{\today}

\begin{abstract}
\end{abstract}
\maketitle

{\bf When the spin-orbit coupling generates a band inversion in a
narrow-bandgap semiconductor such as Sb$_x$Bi$_{1-x}$ or Bi$_2$Se$_3$
the resulting system becomes a strong topological insulator
(STI)\cite{moore1,hasan1}. A key defining property of a STI are its
topologically protected metallic surface states. These are immune to
the effects of non-magnetic disorder and form a basis for numerous
theoretically predicted exotic phenomena\cite{fu3,qi1,essin1,seradjeh1,qi2}
as well as proposed practical applications\cite{nagaosa1,garate1}.
Disorder, ubiquitously present in solids, is normally detrimental to
the stability of ordered states of matter.  In this letter we demonstrate that not
only is STI robust to disorder but, remarkably, under certain
conditions disorder can become fundamentally responsible for its
existence. We show that disorder, when sufficiently strong, can
transform an ordinary metal with strong spin-orbit coupling into a
strong topological `Anderson' insulator, a new topological phase of
quantum matter in three dimensions.
}

Disorder is well known to play a fundamental role in low-dimensional
electronic systems, leading to electron localization and consequent
insulating behavior in the time-reversal invariant systems\cite{lee1}.
Disorder also underlies much of the phenomenology of the integer
quantum Hall effect\cite{evers1}. Recently, in a remarkable
development, it has been noted first by numerical
simulations\cite{jain1} and shortly thereafter by analytical
studies\cite{beenakker1}, that a phase similar to the two dimensional
topological insulator (also known as the quantum spin-Hall
insulator\cite{Zhang2,Konig}) can be brought about by introducing
non-magnetic disorder into a 2D metal with strong spin orbit
coupling. This new 2D topological phase, referred to as topological
Anderson insulator (TAI), has a disordered insulating bulk with
topologically protected gapless edge states that give rise to
precisely {\em quantized} conductance $e^2/h$ per edge. In TAI,
remarkably, conductance quantization owes its very existence to
disorder.

A question naturally arises whether such behavior can occur in three
spatial dimensions. More precisely, one may inquire whether an
inherently 3D topological phase analogous to the {\em strong}
topological insulator\cite{mele1,moore2,roy1} (STI) could be reached
by disordering a clean system that is initially in a topologically
trivial phase. This is a nontrivial question because just as the 3D STI
cannot be reduced to the set of 2D topological insulators, the
existence of a 3D `strong' TAI presumably cannot be deduced from
the physics of 2D TAI. Below, we show the answer to the above question
to be affirmative. Employing a combination of analytical and numerical
methods we construct an explicit example of a disorder-induced
topological phase in three dimensions with physical properties
analogous to those of the strong topological insulator. We propose to
call this new phase a `strong topological Anderson insulator'
(STAI). We argue that some of the topologically trivial compounds with
strong spin-orbit coupling discussed in the recent literature, such as
e.g.\ Sb$_2$Se$_3$, could become STAI upon introducing disorder.  In
other compounds that already are STIs in their clean form, disorder
can reinforce this behavior by rendering the bulk truly insulating. We
note that the authors of Ref.\ [\onlinecite{beenakker1}] anticipated the
existence of a 3D disorder-induced topological phase.

To study the emergence of the STAI we consider a variant of a model describing
 itinerant electrons with spin-orbit coupling  on a cubic lattice discussed
extensively in the recent literature\cite{qi1,hosur1,rosenberg1}. It has four
 electron states per lattice site $\br_j$, compactly denoted as $\Psi_j=(\psi_{1j},\psi_{2j},\psi_{3j},\psi_{4j})^T$, and a
momentum space Hamiltonian $H_0=\sum_\bk\Psi_\bk^\dagger{\cal
H}_\bk\Psi_\bk$ with
$\Psi_\bk$ the Fourier transform of $\Psi_j$,
\begin{equation}\label{hk} {\cal
H}_\bk=\sum_{\mu=0}^3d_\mu(\bk)\Gamma_{\mu}+d_4(\bk)\id,
\end{equation}
and $d_0(\bk)=\epsilon-2t\sum_i\cos{k_i}$,
$d_i(\bk)=-2\lambda\sin{k_i}$ $(i=1,2,3)$ and
$d_4(\bk)=2\gamma\sum_i(1-\cos{k_i})$. Here $\Gamma_\mu$ are $4\times 4$
Dirac matrices in combined orbital and spin space, satisfying the
canonical anticommutation relation
$\{\Gamma_\mu,\Gamma_\nu\}=2\delta_{\mu\nu}$. The system defined by
$H_0$ is invariant under time-reversal and spatial inversion. In the
following we take two alternate points of view regarding Hamiltonian
(\ref{hk}): (i) we view it as a simple toy model conveniently
describing both topological and ordinary phases of non-interacting
electrons in 3D and, (ii) we regard it as a lattice regularization of
the effective low-energy Hamiltonian describing the physics of
insulators in the Bi$_2$Se$_3$ family\cite{zhang1,xia1,shen1}. In the
latter interpretation $\Psi_j$ labels the manifold of four active
orbitals
$(|P1_z^+,\uparrow\rangle,|P2_z^-,\uparrow\rangle,|P1_z^+,\downarrow\rangle,|P2_z^-,\downarrow\rangle)$. In
addition, parameters $t, \lambda$ and $\gamma$ in these materials show
uniaxial anisotropy. We shall ignore this feature for now but return to it
in our discussion of the physical realization of STAI.

The energy spectrum of $H_0$ has two doubly degenerate bands,
\begin{equation}\label{ek}
E_\bk=d_4(\bk)\pm\sqrt{\sum_\mu d_\mu^2(\bk)}.
\end{equation}
At half filling, depending on the values of the parameters $\epsilon$,
$t$, $\lambda$, and $\gamma$ the system can be a metal, a trivial
insulator, as well as a strong and weak topological
insulator\cite{qi1,rosenberg1}. Below, we focus on the topological
phase transition between the ordinary insulator characterized by the
$Z_2$ invariant\cite{mele1,moore2} (0;000) and the (1;000) STI
phase. In the clean system modeled by Hamiltonian (\ref{hk}) this
transition occurs due to the band inversion at the ${\bm \Gamma}$-point of
the Brillouin zone and can be driven by tuning parameter $\epsilon$
through a critical value $\epsilon_c=6t$. We take $t$, $\lambda$ and
$\gamma$ positive here and in what follows. We note that it is
exactly this physics that underlies the STI behavior in the
Bi$_2$Se$_3$ family of materials\cite{zhang1,xia1,shen1}. Our main
result is the finding that a similar transition can be effected by
introducing non-magnetic disorder into a system that is topologically trivial in
its clean form.

To simulate the effects of disorder we consider a Hamiltonian of the form
\begin{equation}\label{hdis} 
H=H_0+\sum_jU_j\Psi^\dagger_j\Psi_j,
\end{equation}
where $H_0$ is the disorder-free Hamiltonian discussed above and $U_j$
is a random on-site potential uniformly distributed in the range
$(-U_0/2,U_0/2)$. Following the discussion in Ref.\
[\onlinecite{beenakker1}] we start by treating the disorder within 
the self-consistent Born approximation (SCBA). The
disorder-averaged electron propagator
$g(\omega,\bk)=(\omega+i\delta+E_F-{\cal H}_\bk-\Sigma_\bk)^{-1}$ is
then given in terms of disorder self-energy $\Sigma_\bk$, subject to
the self-consistent equation
\begin{equation}\label{sig} \Sigma_\bk={U_0^2\over 12}\sum_{\bk\in{\rm
BZ}}\left(E_F+i\delta-{\cal H}_\bk-\Sigma_\bk\right)^{-1}.
\end{equation}
Here $E_F$ refers to the Fermi energy and $\delta$ is a positive
infinitesimal. The factor of 12 arises from the variance $\langle
U^2\rangle=U_0^2/12$. General symmetry consideration restrict the form
of the self-energy to
$\Sigma_\bk=\sum_\mu\Gamma_\mu\Sigma_\mu(\bk)+\id\Sigma_4(\bk)$. In
addition, for point-like disorder $\Sigma_\bk$ is momentum
independent. This forces $\Sigma_i(\bk)$ to vanish for $i=1,2,3$ as
any non-zero value would signal spontaneous time-reversal symmetry
breaking caused by non-magnetic impurities.

The two non-vanishing components of the self-energy, $\Sigma_0$ and
$\Sigma_4$, can be viewed as disorder-induced renormalizations of the
`topological mass' $m\equiv d_0(\bk=0)=\epsilon-6t$ and the Fermi
energy $E_F$, respectively.  Specifically, it is easy to see that
within the SCBA the disorder-averaged system is described by the same
propagator $g_0(\omega,\bk)=(\omega+i\delta+E_F-{\cal H}_\bk)^{-1}$as the clean system but with parameters $m$ and $E_F$
replaced as $m\to \bar{m}=m+\Sigma_0$ and $E_F\to
\bar{E}_F=E_F-\Sigma_4$. We note that $\Sigma_0$ and $\Sigma_4$ are
generally complex-valued, giving $\bar{m}$ and $\bar{E}_F$ both real
and imaginary parts. The latter reflect the quasiparticle lifetime
broadening due to disorder.

These considerations underlie the physical picture behind the
emergence of STAI. Starting from a clean ordinary insulator
(characterized by $m>0$) disorder can induce a band inversion by
driving the real part of the renormalized mass $\bar{m}$
negative. According to the standard classification of topological
insulators\cite{Fu2} a band inversion at an odd number of
time-reversal invariant momenta changes the $Z_2$ class of the
material. If the renormalized Fermi energy $\bar{E}_F$ lies within the
gap the resulting effective medium is a STI.

It is easy to obtain the self-consistent equations for $\bar{m}$ and
$\bar{E}_F$ from Eq.\ (\ref{sig}). These read
\begin{eqnarray} \bar{m}&=&m-{U_0^2\over 12}\sum_\bk{\bar{m}+tc_\bk
\over D_\bk},
 \label{m1}\\ \bar{E}_F&=&E_F+{U_0^2\over 12}\sum_\bk{\bar{E}_F-\gamma
c_\bk \over D_\bk},
\label{e1}
\end{eqnarray}
with $D_\bk= \lambda^2s^2_\bk+(\bar{m}+tc_\bk)^2-(\bar{E}_F-\gamma
c_\bk)^2+i\delta$ and $s_\bk^2=4\sum_i\sin^2{k_i}$,
$c_\bk=2\sum_i(1-\cos{k_i})$. We shall present the full solution to
these below. To gain some insight into the underlying physics it is
useful to first study the approximate solution valid at weak disorder (small
$U_0$) obtained by replacing $\bar{m}$ and $\bar{E}_F$ on the right
hand side by their respective bare values. One then obtains
\begin{eqnarray} \bar{m}&\simeq&m-{U_0^2\over 24\pi}{t \over
t^2-\gamma^2},
\label{m2}\\ \bar{E}_F&\simeq&E_F+{U_0^2\over 24\pi}{\gamma \over
t^2-\gamma^2},
\label{e2}
\end{eqnarray}
where we kept only the leading divergent terms after expanding the
integrand around the ${\bm \Gamma}$ point. We observe that for $t>\gamma$
disorder indeed renormalizes the mass term downward; for
$U_0>U_c\simeq [24\pi m(t^2-\gamma^2)/t]^{1/2}$ band inversion occurs
and one expects the system to become a STAI. We demonstrate below that
this conclusion remains valid when Eqs. (\ref{m1},\ref{e1}) are solved
self-consistently, as well as in the numerical simulation of Hamiltonian (\ref{hdis}).
\begin{figure*}
\includegraphics[width = 14.0cm]{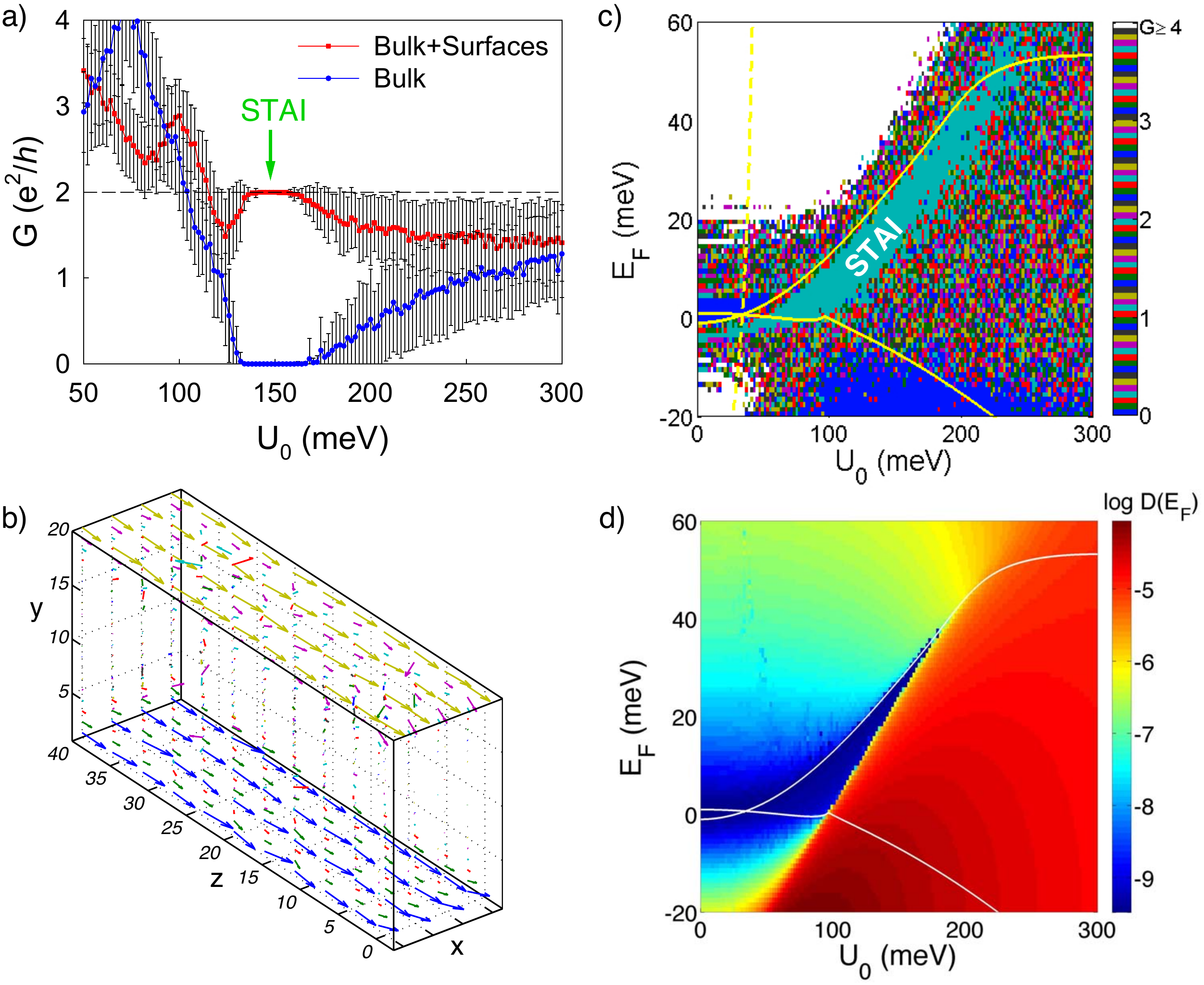}
\caption{{\bf Conductance in disorder-induced phases. }
a) Conductance $G$ of a rectangular wire with dimensions $4\times
20\times 40$ as a function of disorder strength $U_0$ for periodic boundary conditions along $x$ and periodic 
(open) boundary conditions along $y$ as distinguished by blue (red) symbols,
with $E_F=20$meV. Error bars reflect the conductance fluctuations in
the ensemble of 100 independent disorder realizations.  b) Electrical
current distribution in the STAI wire averaged over 100 independent disorder realization for $E_F=20$meV and $U_0=150$meV. We use open boundary conditions along $y$ and periodic along $x$ direction. Arrows representing the local current density in different layers along the $y$-direction have been
color-coded for clarity.   c)
False color plot of conductance as a function of disorder strength
$U_0$ and the Fermi energy $E_F$. Each data point corresponds to a
single realization of the disorder potential. The color scale is chosen to emphasize the effect of fluctuations in $G$. The dashed line marks
the band inversion boundary defined as ${\rm Re}(\bar{m})=0$ based on
the SCBA. Solid lines represent the SCBA phase boundaries separating a
band insulator and a metal defined by $|{\rm Re}(\bar{E}_F)|=|{\rm
Re}(\bar{m})|$.  d) Density of states at the Fermi energy $D(E_F)$ calculated using SCBA.
 In all four panels we use parameters $\epsilon=145$meV, $t=24$meV,
$\lambda=20$meV, and $\gamma=16$meV, corresponding to $m=1$meV.
}
\label{fig1}
\end{figure*}

The expression for the critical disorder strength $U_c$ derived above
suggests that a significant amount of disorder might be necessary to
effect the band inversion in a real material. Since the Born
approximation is expected to hold in the limit of weak disorder it is
important to establish the stability of STAI phase by means of a
complementary method. To this end we carry out exact numerical
diagonalization studies of our model Hamiltonian
(\ref{hdis}). Generalizing the approach of Refs.\
[\onlinecite{jain1,beenakker1,jiang1}] we study the emergence of
topologically protected gapless surface states that are the defining
feature of the strong topological insulator in 3D. In a disordered
system it is not sufficient to establish the existence of gapless
modes as these could be localized in space. Instead, one must seek
extended states capable of carrying electrical current across the
macroscopic sample. With this in mind we employ the recursive Green's
function method\cite{bruno1} to evaluate the conductance $G$ of a length-$L$ wire
with a rectangular cross-section $W_x\times W_y$ using the
Landauer-B\"uttiker formalism\cite{landauer1,buttiker1}. For
simplicity and concreteness, and to make contact with the previous
works on 2D systems\cite{jain1,beenakker1,jiang1}, we use model
parameters listed in the caption of Fig.\ \ref{fig1}, with values
close to those describing HgTe/CdTe quantum wells\cite{Zhang2,Konig}.

Figure \ref{fig1}a shows conductance $G$ as a function of disorder
strength $U_0$ in a wire with $W_x=4$ and $W_y=20$. For weak disorder
the wire shows conductance characteristic of a disordered metal with significant fluctuations reflecting different
realizations of the disorder potential $U_i$. Above $U_0\simeq 130$meV the
bulk conductance (measured with periodic boundary conditions along $x$
and $y$) drops to zero, indicating a disorder-induced insulating
behavior in the bulk that persists up to $U_0\simeq 170$meV. If we
change periodic boundary conditions to {\em open} along the
$y$-direction a very different picture emerges. For $U_0$ in the range
showing bulk insulating behavior the conductance is now pinned to
the non-zero value $2e^2/h$ with no observable fluctuations. We
attribute this to the ballistic transport of topologically protected
surface states.  Figure \ref{fig1}b confirms that the current flows
near the surface only. It furthermore shows that the current density
is equally distributed among the two surfaces; each surface forms an
independent conduction channel contributing one quantum $e^2/h$ to the
total conductance $G$ of the sample. This is the disorder-induced
topological phase mentioned above.

How can one be sure that this is a genuinely 3D topological phase
analogous to STI and not merely a result of 2D physics already
discussed in Refs.\ [\onlinecite{jain1,beenakker1}]? First, we note
that our model Hamiltonian is completely isotropic in 3D space so it
is unlikely that the observed behavior would originate from the
edge states of a set of 2D layers. We checked that the conductance
remains quantized at $2e^2/h$ when we change the number of layers $W_x$  and
there is no even/odd effect that one would expect in a layered
system. The
odd number of 1D conductance channels per surface reflects the odd
number of gapless states per surface of a STAI, confined to a 1D
geometry imposed by the finite width of the wire in the
$x$-direction. When we impose anti-periodic boundary conditions for
the electron wavefunctions along the $x$-direction the conductance
drops to zero in accord with the expectation for the surface state in
a 3D topological phase. 
\begin{figure}
\includegraphics[width = 7.5cm]{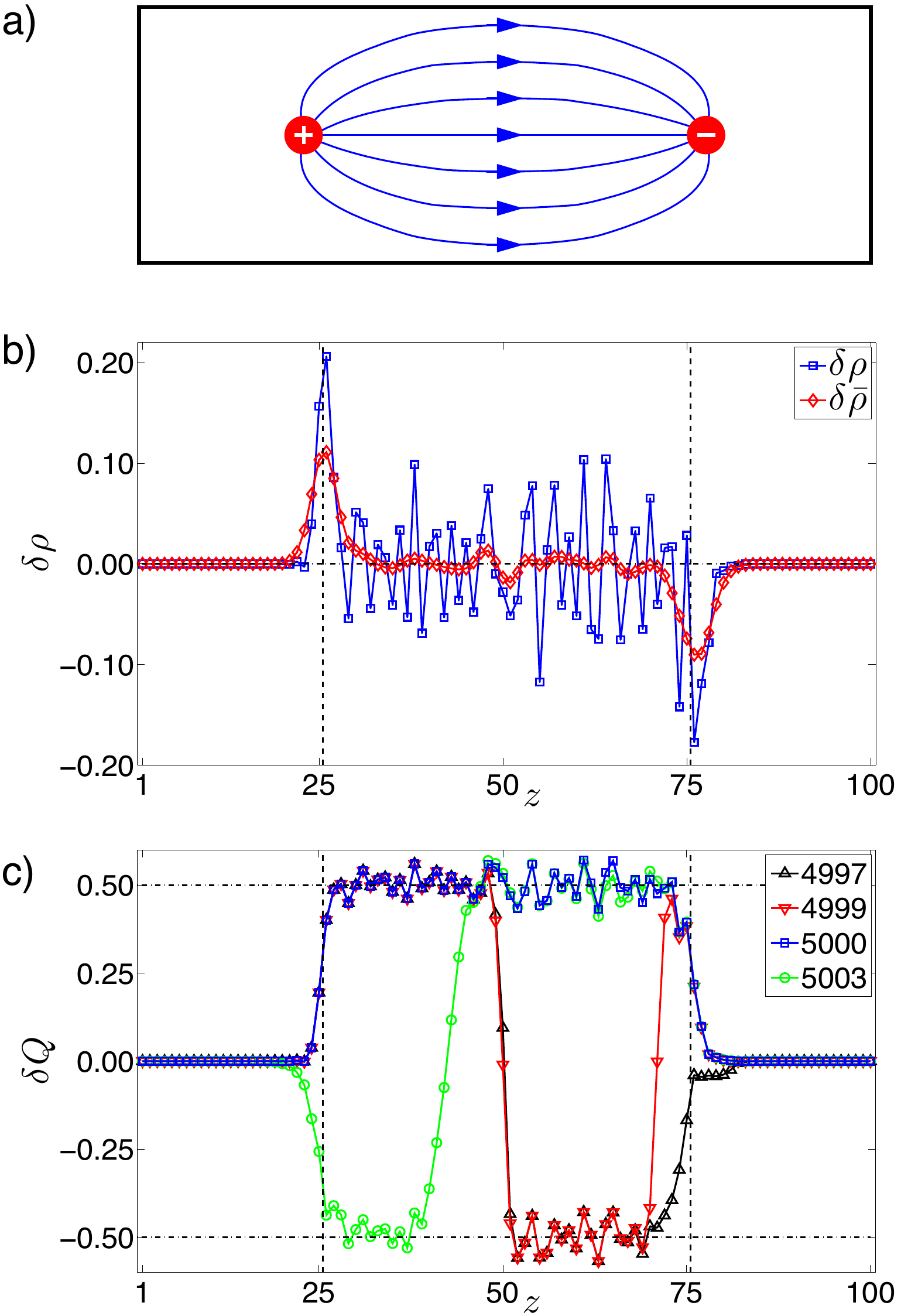}
\caption{{\bf Witten effect in the strong topological Anderson insulator. }
a) A prism-shaped sample with monopole (+), anti-monopole ($-$), and
field lines indicated. The asymmetric field distribution allows us to
use periodic boundary conditions in all directions and thus avoid
difficulties associated with the surface states. 
b) Electric charge density $\delta\rho(z)=\sum_{x,y}[\rho(\br)-
\rho_0(\br)]$ induced by the monopole/anti-monopole pair in a $5\times 5\times 100$
sample at half filling. Here $\rho$ and $\rho_0$ represent the charge
density with and without the pair, respectively. A smoothed charge density function
$\delta\bar\rho(z)$, obtained by convolving $\delta\rho(z)$ with a
Gaussian of width $\sigma=1.5$, is also plotted to emphasize charge bound to the monopole.
c) Integrated charge $\delta Q(z)=\sum_{z'\leq z}\delta\rho(z')$ in
units of $e$ for  $E_F$ in the range $19-25$meV, corresponding to the STAI phase. The number of filled  electron states is indicated in the legend (5000
represents the half-filling). The steps of magnitude $\pm e/2$ at the monopole/anti-monopole
locations show the expected localized fractional charge due to the
Witten effect. Steps with magnitude $\pm e$ located
elsewhere are due to the shift of some bound states in applied
magnetic field and can be viewed as a finite-size effect which should
diminish for a system with a larger cross section. In both panels we use
$U_0=150$meV and other parameters as in Fig.\ 1. 
}
\label{fig2}
\end{figure}

To further confirm the 3D nature of the observed topological phase we
probed for the Witten effect\cite{witten1} in our model
STAI. According to
Refs.\ [\onlinecite{qi1,essin1}] the effective electromagnetic Lagrangian of a 3D strong topological insulator contains an unusual `axion' term
$\sim\theta\bE\cdot\bB$  with $\theta=\pi$. According to Witten\cite{witten1}
a  magnetic monopole
inserted into a medium with non-zero $\theta$ binds electric charge
$-e(\theta/2\pi+n)$ with $n$ integer. Using numerical methods described in Ref.\ [\onlinecite{rosenberg1}] we 
measured the induced fractional charge in a configuration containing a
monopole and an anti-monopole depicted
in Fig.\ \ref{fig2}a. Our results presented in Fig.\ \ref{fig2}b,c
clearly indicate fractional charge $\pm e/2$ bound to the monopole,
confirming the expected value of $\theta=\pi$. This result lends
additional support to our identification of STAI as a genuinely 3D
topological phase characterized by the bulk axion term.

We now turn to the phase diagram of our model. Specifically,
we wish to map out the locus of points in the space of parameters
$(E_F,U_0)$ that gives rise to STAI behavior. In a real material these
parameters can be tuned, at least in principle, by adjusting the
chemical composition and disorder content.  Because of the 3D nature
of our system and the resulting large size of the Hamiltonian matrix
that must be diagonalized, we were able to consider only a single
realization of the disorder potential $U_i$ at each point of the
$(E_F,U_0)$ phase diagram. Nevertheless this turns out to be
sufficient for determining the location of STAI phase to a good
accuracy. Our method relies on the fact that, as seen in Fig.\ \ref{fig1}a,
the conductance $G$ shows no observable fluctuations in the STAI phase but
fluctuates significantly elsewhere. Figure \ref{fig1}c displays $G$ 
in a fashion that is designed to amplify the effect of
fluctuations. The locus of STAI phase is clearly visible as the region
with $G=2e^2/h$ and no discernible conductance fluctuations.

The weak-disorder boundary of the numerically determined STAI phase
coincides with the metal-insulator transition deduced from the SCBA
Eqs.\ (\ref{m2},\ref{e2}). As in 2D, the weak-disorder STAI phase
boundary marks the crossing of a band edge rather than a mobility
edge\cite{beenakker1}. The strong-disorder phase boundary is more
interesting. According to SCBA (Fig.\ \ref{fig1}d) states appear at
the Fermi level beyond certain disorder strength $U^*(E_F)$ due to
lifetime broadening. The band edge becomes ill-defined here as the
self energy acquires a large imaginary part. Comparison to Fig.\
\ref{fig1}c shows that the numerically determined STAI phase extends well
beyond $U^*(E_F)$. The bulk electron states near the Fermi level do
not contribute to conduction in this regime and must therefore be
localized. It is tempting to identify this region as the `true'
topological Anderson insulator where electron localization plays the
key role.

Although discussed here in the framework of a concrete model we expect
the emergence of disorder-induced topological phases to be quite
generic in three spatial dimensions. We have verified by an explicit
calculation that including uniaxial anisotropy characteristic of the
materials in the Bi$_2$Se$_3$ family\cite{zhang1,xia1,shen1} leads to
qualitatively similar behavior to that displayed in Fig.\
\ref{fig1}. This suggests that clean Sb$_2$Se$_3$, predicted to be a
trivial insulator\cite{zhang1} (but nevertheless close to the
topological phase), could become STAI upon introducing non-magnetic
disorder. Whether or not disorder can effect a band inversion in
Sb$_2$Se$_3$ will depend crucially on the magnitude of its native
bandgap. Our simulations suggest that when the bandgap size becomes
larger than $\sim 20$meV the amount of disorder required to produce a
band inversion is so large that the electron states become localized
before STAI phase can be reached. From this point of view the best
prospects for experimental realization of STAI physics lie with
materials that are very-small or zero bandgap semiconductors with
strong spin-orbit coupling. Although bulk HgTe as well as the recently
discussed Heusler compounds\cite{chadov1,lin1} exhibit this type of
behavior their low-energy physics is not well described by our
four-band model and further theoretical work is needed to determine whether
disorder could drive the transition into the topological phase.

{\em Acknowledgments ---}
The authors acknowledge illuminating discussions with I.~Garate,
A.~Kitaev, J.E.~Moore, A.~Vishwanath, C.~Weeks and S.-C.~Zhang.
The work was supported in part by NSERC, CIfAR (MF), China Scholarship Council (HMG), the
Packard Foundation, and the Research Corporation (GR).




\begin{thebibliography}{10}
\bibitem{moore1} J.E.~Moore, Nature {\bf 464}, 194 (2010).
\bibitem{hasan1} M.Z.~Hasan, C.L.~Kane, arXiv:1002.3895.

\bibitem{fu3} L.~Fu and C.L.~Kane, \prl {\bf 100}, 096407 (2008).

\bibitem{qi1} 
X.-L.~Qi, T.~Hughes, and S.-C.~Zhang, \prb {\bf 78}, 195424 (2008).

\bibitem{essin1} A.M.~Essin, J.E.~Moore, D.~Vanderbilt, \prl {\bf 102}, 146805 (2009).

\bibitem{seradjeh1} B.~Seradjeh, J.E.~Moore, and M.~Franz, \prl {\bf 103}, 066402 (2009).

\bibitem{qi2} X. L.~Qi {\em et al.}, Science {\bf 323}, 1184 (2009).

\bibitem{nagaosa1} T.~Yokoyama, Y.~Tanaka, and N.~Nagaosa, \prb {\bf 81}, 121401(R) (2010).

\bibitem{garate1} I.~Garate and M.~Franz, \prl {\bf 104}, 146802 (2010).

\bibitem{lee1} P.A.~Lee and T.V.~Ramakrishnan,  Rev.~Mod.~Phys.~{\bf 57}, 287 (1985). 

\bibitem{evers1} F.~Evers and A.D.~Mirlin, Rev.~Mod.~Phys.~{\bf 80}, 1355–1417 (2008).

\bibitem{jain1} J.~Liu, R.-L.~Chu, J.K.~Jain and S.-Q.~Shen, \prl {\bf 102}, 136806 (2009).

\bibitem{beenakker1} C.W.~Groth, M.~Wimmer, A.R.~Akhmerov, J.~Tworzydlo and C.W.J.~Beenakker, \prl {\bf 103}, 196805 (2009).

\bibitem{Zhang2} B.A.~Bernevig, T.L.~Hughes, and S.-C.~Zhang,
  Science {\bf 314}, 1757 (2006).
  
 \bibitem{Konig}
M.~K\"{o}nig {\em et al.}, Science {\bf 318}, 766 (2007).

\bibitem{mele1} 
L.~Fu, C.~L.~Kane, and E.~J.~Mele, \prl {\bf 98} 106803 (2007).

\bibitem{moore2} J.~E.~Moore and L.~Balents, \prb {\bf 75} 121306(R) (2007). 

\bibitem{roy1} R.~Roy, \prb {\bf 79}, 195322 (2009).

\bibitem{hosur1} P.~Hosur, S.~Ryu, and A.~Vishwanath, \prb {\bf 81}, 045120 (2010).

\bibitem{rosenberg1} G.~Rosenberg and M.~Franz, \prb (in press, arXiv:1001.3179).

\bibitem{zhang1} H.~Zhang, C.-X.~Liu, X.-L.~Qi, X.~Dai, Z.~Fang, and
  S.-C.~Zhang, Nature Phys. {\bf 5}, 438 (2009).

\bibitem{xia1} Xia, Y. {\em et al.}  
Nature Phys.~{\bf 5}, 398–402 (2009).

\bibitem{shen1} Chen, Y.L.~{\em et al.}  
Science {\bf 325}, 178–181 (2009).


\bibitem{Fu2} L.~Fu and C.~L.~Kane, \prb {\bf 76}, 045302 (2007).

\bibitem{jiang1} H.~Jiang, L.~Wang, Q.-F.~Sun and X.C.~Xie, \prb {\bf 80}, 165316 (2009).

\bibitem{bruno1} We use the variant of the method described by G.~Metalidis and P.~Bruno, \prb {\bf 72}, 235304 (2005).

\bibitem{landauer1} R.~Landauer, Philos.\ Mag. {\bf 21}, 863 (1970).

\bibitem{buttiker1} M.~B\"uttiker, \prb {\bf 38}, 9375 (1988).

\bibitem{witten1} E.~Witten, Phys.\ Lett.\ B {\bf 86}, 283 (1979).


\bibitem{chadov1} S.~Chadov {\em et al.}, Nature Mat.~(in press, arXiv:1003.0193).

\bibitem{lin1} H.~Lin {\em et al.}, Nature Mat. (in press, arXiv:1003.0155).




\end{thebibliography}
\end{document}